\documentclass{PoS}
\usepackage{slashed}    
\title{Calculation of the $K_L - K_S$ mass difference for physical quark masses}

\ShortTitle{$\Delta m_K$ for physical quark masses}

\author{\speaker{Bigeng Wang}\thanks{This work was partially supported by US DOE grant
\#DE-SC0011941 and
used computer time provided by the Innovative and Novel Computational Impact
on Theory and Experiment (INCITE) program. This research used resources of
the Argonne Leadership Computing Facility, which is a DOE Office of Science
User Facility supported under Contract DE-AC02-06CH11357.}\\
        Department of Physics, Columbia University, New York, NY 10027, USA\\
        E-mail: \email{bw2482@columbia.edu}}

\abstract{In this article, I will present the status of our calculation of the difference between the masses of the long- and short-lived neutral K mesons, $\Delta m_K$ predicted by the Standard Model. This calculation is performed on an ensemble of 152, $64^3 \times 128$ gauge configurations with an inverse lattice spacing of 2.36 GeV and physical quark masses. The results from different methods of analysis and our progress toward obtaining a final result will be discussed.}

\FullConference{The 37th Annual International Symposium on Lattice Field Theory - LATTICE2019\\
		16-22 June, 2019\\
		Wuhan, China.}
\begin{document}

\section{Introduction}

The mass difference between $K_L$ and $K_S$ is a quantity related to $\Delta S = 2$ weak interaction and very sensitive to new physics beyond the Standard Model. Due to the GIM mechanism, this quantity receives contribution mainly from charm quark scale where the accuracy of QCD perturbative calculation is limited by the strong coupling \cite{gorbahn}. However, the non-perturbative calculation using lattice QCD with a sufficiently small lattice spacing should provide results with controlled systematic errors. Following the RBC-UKQCD collaborations' first full calculation with unphysical kinematics\cite{jianglei_2}, our most recent calculation is performed on a $64^3 \times 128$ lattice with physical masses on 152 configurations. In this article, preliminary results, methods used for reducing statistical errors and discussion of systematic errors are presented.

\section{$\Delta m_K$ and GIM mechanism}
The K meson mixing through $\Delta S = 2$ weak interaction within the Standard Model is described by diagrams of the sort shown below in Figure \ref{fig:box_diag}.
\begin{figure}
    \centering
    \includegraphics[width=0.5\textwidth]{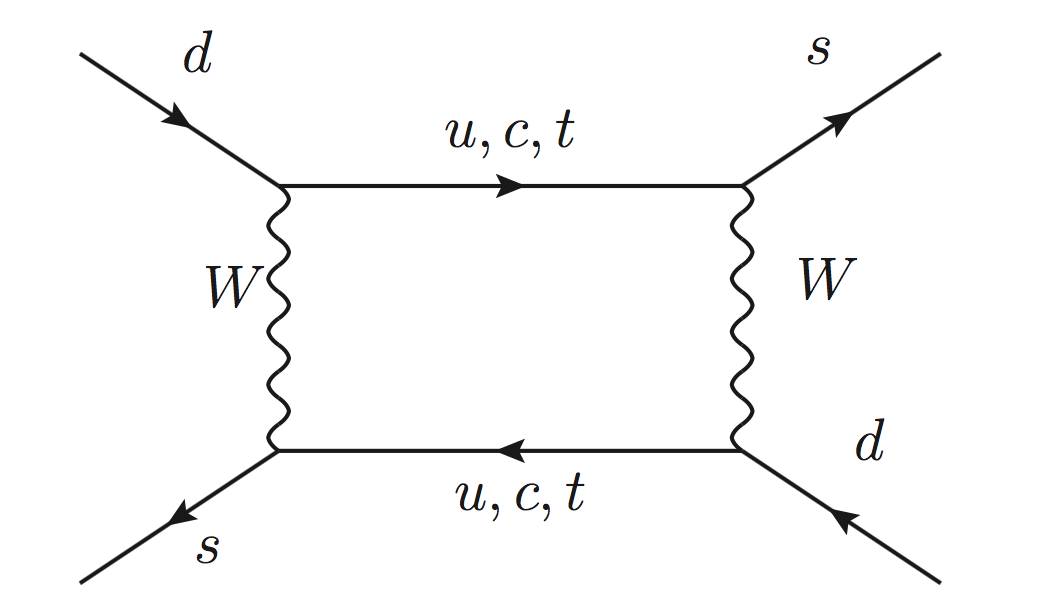}
    \caption{The box diagram contributing to kaon mixing in the Standard Model.}
    \label{fig:box_diag}
\end{figure}

With only up quark propagators , the loop integral yields quadratic ultraviolet divergence. On the lattice, this is regulated by the finite lattice spacing $a$, i.e. an ultraviolet cutoff at energy $\Lambda \sim a^{-1}$:
\begin{equation}
    \int_{m_u}^{\Lambda} d^4p \gamma^{\mu}(1-\gamma_5) \frac{\slashed{p}-m_u}{\slashed{p}^2+m_u^2}\gamma^{\nu}(1-\gamma_5)\frac{\slashed{p}-m_u}{\slashed{p}^2+m_u^2}
\end{equation}\\
However, for the four-flavor case, the GIM mechanism in K meson mixing leads to the difference of up and charm quark propagators appearing within the loop. In addition, specific to the $\Delta m_K$ calculation, due to the left-left spin structure of the two weak operators involved, the ultraviolet part of the loop integration becomes:
\begin{equation}
        \int_{m_c}^{a^{-1}} d^4p \gamma^{\mu}(1-\gamma_5) (\frac{\slashed{p}(m_c^2-m_u^2)}{(\slashed{p}^2+m_u^2)(\slashed{p}^2+m_c^2)})\gamma^{\nu}(1-\gamma_5)(\frac{\slashed{p}(m_c^2-m_u^2)}{(\slashed{p}^2+m_u^2)(\slashed{p}^2+m_c^2)}).
\end{equation}
As a result both quadratic and logarithmic divergences are removed and this ultraviolet contribution to $\Delta m_K$ becomes:
\begin{equation}
    \sim m_c^4(\frac{1}{m_c^2}-\frac{1}{a^{-2}})\sim m_c^2(1 + \mathcal{O}(m_ca)^2).
\end{equation}
From this we could conclude:
\begin{itemize}
    \item There's no ultraviolet divergence and the physics scale related to $\Delta m_K$ is at the charm mass.
    \item The effect of ultraviolet cutoff arising on the loop momentum integral, the finite lattice spacing $a$, is the same size as other finite lattice spacing effect from the charm mass. Thus there is no need for local short distance correction beyond the bi-local $\Delta S = 1$ operators used here.
\end{itemize}
Previous tests on $16^3\times32$ lattice have shown behaviours consistent with above conclusions \cite{jianglei_pos}:
\begin{itemize}
    \item In the 3-flavor calculation of $\Delta m_K$ from operator product $Q_1Q_1$, a quadratic dependence on the inverse of an artificially introduced cutoff radius is observed. 
    \item In the 4-flavor calculation with the GIM mechanism, this dependence on the cut off radius $R$ disappears for small $R$.
\end{itemize}
As a result, our $64^3\times 128$ physical mass lattice calculation needs only the usual multiplicative renormalization of the four-quark weak operators.



\section{$\Delta m_K$ on lattice and integrated correlators}
The $K_L-K_S$ mass difference is expressed as:
\begin{equation}
    \label{eqn:dmk}
    \Delta M_K = 2 Re M_{\overline{0}0} = 2 \mathcal{P} \sum_{n} \frac{\langle \bar{K}^0|H_W|n\rangle\langle n|H_W|K^0\rangle}{m_K-E_n},
\end{equation}
where $H_W$ is the $\Delta S =1 $ effective Hamiltonian:
\begin{equation}
\label{eqn:hamiltonian}
    H_W = \frac{G_F}{\sqrt{2}} \sum_{q,q'=u,c} V_{qd}V^*_{q's}
    (C_1Q_1^{qq'}+C_2Q_2^{qq'}).
\end{equation}
Here the ${Q_i^{qq'}}_{i=1,2}$ are current-current operators, defined as:
\begin{equation}
\label{eqn:q1q2}
Q_1^{qq'} = (\bar{s}_i \gamma^{\mu} (1- \gamma_5)d_i)(\bar{q}_j \gamma^{\mu}(1-\gamma_5)q_j'), \quad
Q_2^{qq'} = (\bar{s}_i \gamma^{\mu} (1- \gamma_5)d_j)(\bar{q}_j \gamma^{\mu}(1-\gamma_5)q_i'),
\end{equation}
where $i$ and $j$ are color indices and $V_{q_a q_b}$ are the usual CKM matrix elements and $C_i$ are Wilson coefficients.

We obtain the Wilson coefficients $C_i^{lat}$ for the lattice operators in three steps \cite{NPR} \cite{NPR2}:
\begin{itemize}
        \item 
        Non-perturbative renormalization: Renormalize the lattice in the RI-SMOM renormalization scheme.
        \item
        Perturbation theory: Convert from RI-SMOM to $\overline{MS}$ renormalization.
        \item
        Perturbation theory: Calculate the Wilson coefficients in the $\overline{MS}$ scheme.
\end{itemize}

To evaluate $\Delta m_K$ on an Euclidean-space lattice, we have previously calculated double-integrated correlators \cite{bigeng_18} integrating over the time locations of both weak operators. We could also evaluate the single-integrated correlators \cite{xu}:
    \begin{equation}
    \mathcal{A}^S (T)= \frac{1}{2!} \sum^{t_1+T}_{t_2=t_1-T} 
    \langle 0 |T \{ \bar{K}^0(t_f)H_W(t_2)H_W(t_1)K^0(t_i) \} | 0 \rangle.
    \label{eqn:single_intgrate}
    \end{equation}
    Similar to the double-integrated case, we insert a complete set of intermediate states and get: 
    \begin{equation}
    \mathcal{A}^S (T)=N_K^2 e^{-m_K(t_f-t_i)} \sum_{n} \frac{\langle \bar{K}^0|H_W|n\rangle\langle n|H_W|K^0\rangle}{m_K-E_n} \{-1 +e^{- (E_n - m_K)T} \}.
    \end{equation}

    \begin{figure}
    \centering
    \includegraphics[width=0.8\textwidth]{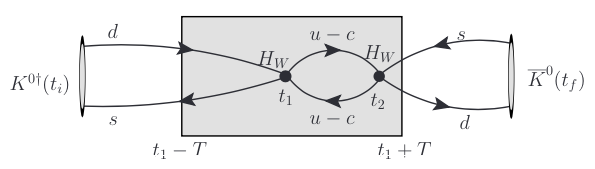}
    \caption{The single integration method on lattice. The shadowed box refers to the region of integration.}
    \label{fig:single_int_method}
\end{figure}
 We obtain $\Delta m_K$ frpm the constant term in Equation \ref{eqn:dmk}. However, to do this the terms which are exponential increasing with increasing $T$ coming from states $|n\rangle$ with $E_n < m_K$ must be removed.
\subsection{$\mathcal{O}(a)$ finite lattice spacing error}
On the lattice, the time integral is replaced by a sum over time slices and this may introduce finite lattice spacing errors $\sim \mathcal{O}(a)$. 
In the double-integration method, this effect is eliminated by the symmetry of the integration. In our single-integration method, after the exponentially growing contribution from states with $E_n < m_K$ has been removed the resulting unintegrated correlator vanishes near the integration limits, and any $\mathcal{O}(a)$ contribution is suppressed.
\subsection{Exponentially growing term subtraction}
In our case of physical quark masses, the $|0\rangle$, $|\pi\pi\rangle_{I=0,2}$, $|\eta\rangle$ and  $|\pi\rangle$ states have energy either smaller or slightly larger than $m_K$ and therefore need to be subtracted. With the freedom of adding the operators $\overline{s}d$ and $\overline{s} \gamma_5 d$ to the weak Hamiltonian with properly chosen coefficients $c_s$ and $c_p$, we are able to remove two of these contributions. Here we choose $c_s$ and $c_p$ to satisfy Equation \ref{eqn:cpcs} so that contributions from the $|0\rangle$ and $|\eta\rangle$ will vanish: 
\begin{equation}
\label{eqn:cpcs}
    \langle 0 | H_W - c_p \bar{s}\gamma_5 d| K^0\rangle=0, \quad
    \langle \eta | H_W - c_s \bar{s}d| K^0\rangle=0.
\end{equation}
As a result, the current-current operators in the original $\Delta S = 1$ effective weak Hamiltonian in Equation \ref{eqn:hamiltonian}  should be modified to be :
\begin{equation}
   Q_i'= Q_i - c_{pi} \bar{s}\gamma_5 d - c_{si} \bar{s}d
\end{equation}
with $c_{pi}$ and $c_{si}$ are calculated on lattice using Equation \ref{eqn:cscp_calc}.\\
\begin{equation}
\label{eqn:cscp_calc}
    c_{si}=\frac{\langle \eta | Q_i | K^0\rangle}{\langle \eta | \overline{s}d | K^0\rangle},  \quad
    c_{pi}=\frac{\langle 0 | Q_i | K^0\rangle}{\langle 0| \overline{s}\gamma_5 d | K^0\rangle}.
\end{equation}

\begin{figure}
    \centering
    \includegraphics[width=0.7\textwidth]{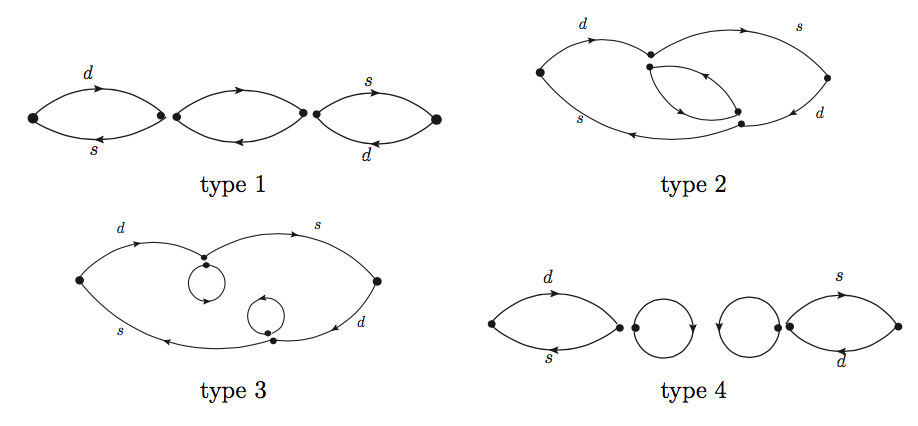}
    \caption{Four types of contractions in the 4-point correlators with $Q_1$ and $Q_2$.}
    \label{fig:contract_q}
\end{figure}

For contractions among $Q_i$, there are four types of diagrams to be evaluated, as shown in Figure \ref{fig:contract_q}. In addition, there are "mixed" diagrams from the contractions between the $\bar{s}d$, $\bar{s}\gamma_5d$ and $Q_i$ operators, having similar topologies to type 3 and type 4 contractions.

\section{Lattice calculation and results}
The calculation was performed on a $64^3 \times 128 \times 12$ lattice with 2+1 flavors of M$\ddot{o}$bius DWF and the Iwasaki gauge action with physical pion mass (136 MeV) and inverse lattice spacing $a^{-1} = 2.36$ GeV. The input parameters are listed in Table \ref{tab:inputs}. Compared to the  results presented in Lattice 2018, we still have in total 152 configurations but now use the single-integration method which yields consistent $\Delta m_K$ results with smaller statistical errors. The results for two-point and and three-point correlators are identical and could be found in the paper of last year\cite{bigeng_18}. Here I only present the results from four-point correlators.

\begin{table}[]
    \centering
    \begin{tabular}{ | c  | c | c | c | c |}
    \hline
   $\beta$ & $am_l$  & $am_h$ & $\alpha=b+c$ & $L_s$\\ \hline
 2.25 & 0.0006203 & 0.02539 & 2.0 &  12 \\ \hline
	
    \end{tabular}
    \caption{Input parameters of the lattice calculation.}
    \label{tab:inputs}
\end{table}
\subsection{Four-point correlators}
In our single-integration method, we subtract the light states before integration and expect the resulting unintegrated correlator to decrease exponentially as the time separation between the two weak operator $\delta \equiv |t_1 - t_2|$ increases. By examining the values of unintegrated correlators, we can identify the range of $\delta$ where the contributions are consistent with zero and therefore avoid including their contributions to statistical errors.

\begin{figure}

    \centering
        \includegraphics[width=0.48 \textwidth]{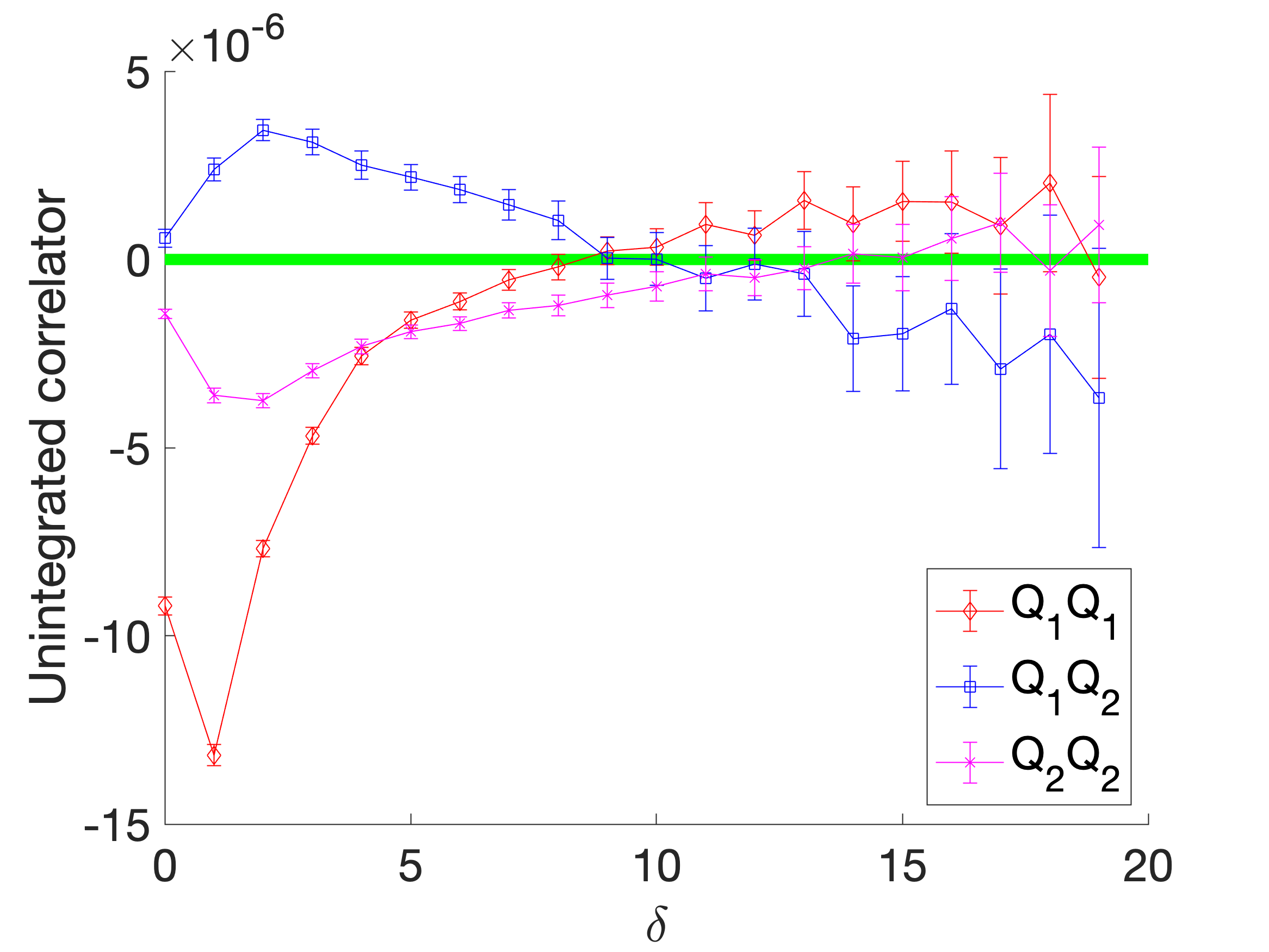}
        \includegraphics[width=0.48 \textwidth]{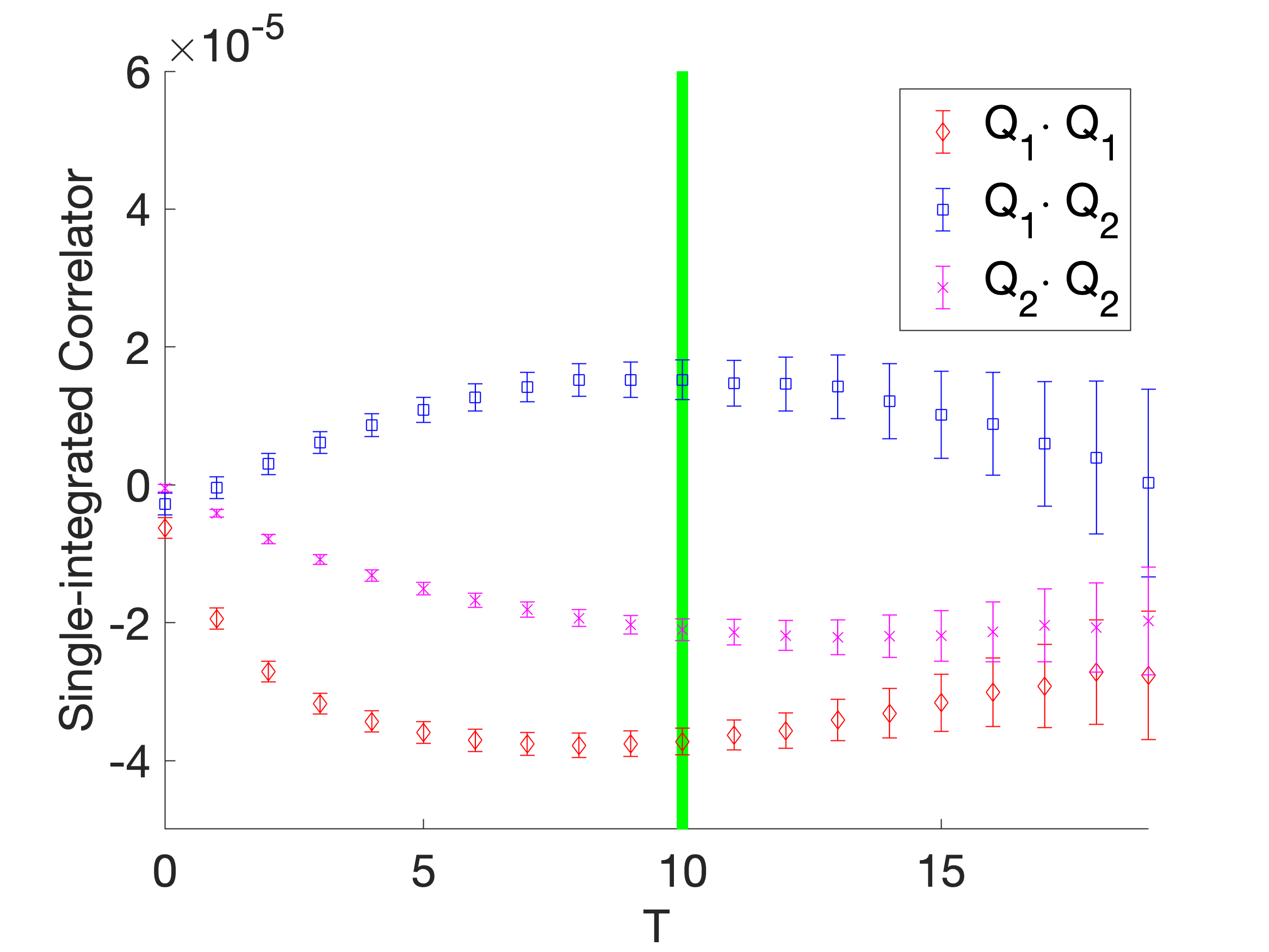}
    \caption{The four-point correlators. The left plot shows the unintegrated correlator obtained from an error-weighted average over all locations of the pair of operators subject to the constraint that neither operator is closer to the single kaon operators than 10 time units. The right plot shows the correlators $\mathcal{A}^S_{ij}(T)$ obtain by integrating the data shown in the left plot over $\delta$. }
    \label{fig:DMK}
\end{figure}

The unintegrated four-point correlators with respect to  $\delta$ are plotted in Figure \ref{fig:DMK} . From the unintegrated correlators ploted, we find for $\delta \textgreater 10$ the values of correlators are zero within uncertainties. Thus we choose the integration upper limit $T = 10$ and obtain $\Delta m_K$ from the single-integrated correlators $\mathcal{A}_{ij}^{S}(T=10)$, where $i,j = 1,2$. The $\Delta m_K$ value extracted are shown in Table \ref{tab:DMK}. Compared to the previously obtained double-integrated value, the new results have smaller statitical errors . 
\begin{table}[!htb]
    \centering
    \begin{tabular}{c|c|c|c} \hline
       Method  & $\Delta m_K$ & $\Delta m_K$(tp1\&2) &$\Delta m_K$(tp3\&4)\\ \hline
         Double-integration & 8.2(1.3) & 8.3(0.6) & 0.1(1.1)  \\ \hline
         Single-integration & 6.90(0.58) & 7.11(0.30) & -0.29(0.49)  \\ \hline
    \end{tabular}
    \caption{Results for $\Delta m_K$ from uncorrelated fits in units of $10^{-12}$ MeV with fitting range 10:20.}
    \label{tab:DMK}
\end{table}
\begin{figure}
    \centering
        \includegraphics[width=0.48 \textwidth]{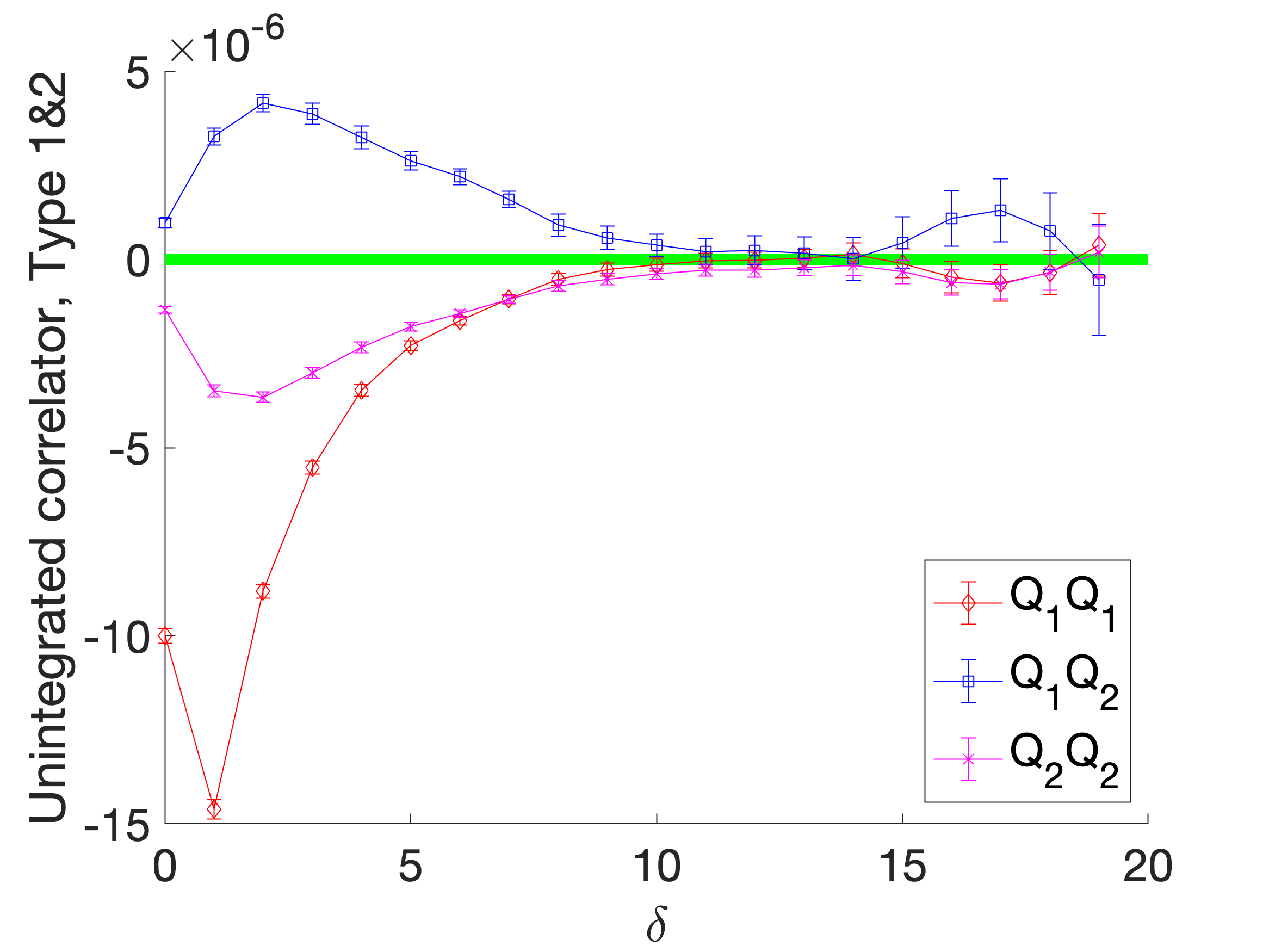}
        \includegraphics[width=0.48 \textwidth]{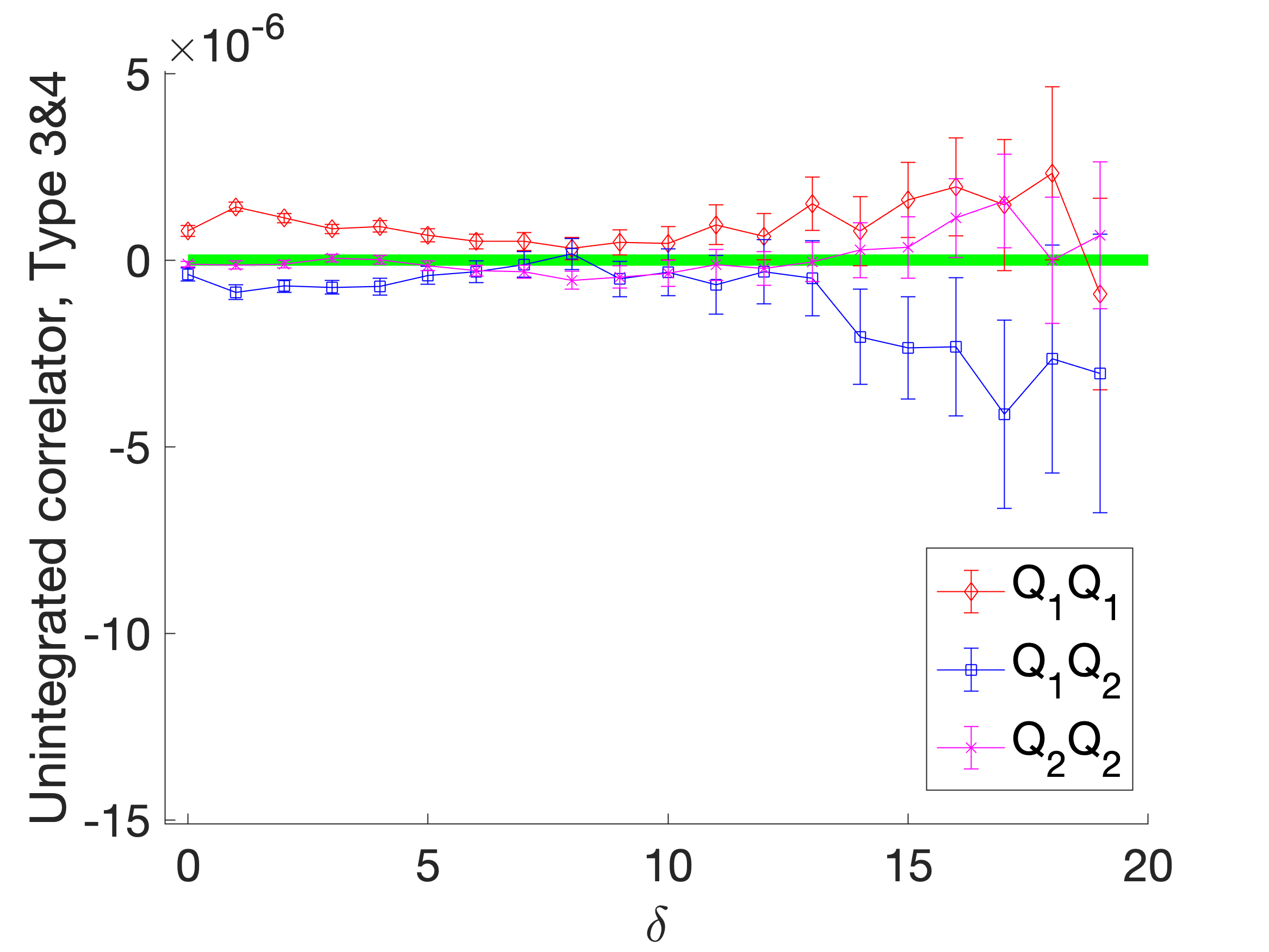}
        
    \caption{Unintegrated correlators from type 1 and 2 diagrams(left) and type 3 and type 4 diagrams(right). }
    \label{fig:DMK_unint_type12_34}
\end{figure}

The unintegrated correlators from the different types of diagrams are ploted in Figure \ref{fig:DMK_unint_type12_34} and corresponding contributions to $\Delta m_K$ are shown in Table \ref{tab:DMK}. The main contribution to $\Delta m_K$ is from type 1 and type 2 diagrams and the contribution from type 3 and 4 having disconnected pieces is zero within uncertainty. This may imply the validity of the OZI rule in the case of physical kinematics in contrast to the previous calculation of $\Delta m_K$ with unphysical kinematics, where contributions from type 3 and 4 diagrams are almost half of the contributions from type 1 and type 2 diagrams with opposite sign \cite{jianglei_2}. 
\section{Systematic errors}
Two potentially important systematic errors come from finite-volume and finite lattice spacing effects. The finite-volume correction to $\Delta m_K$ based on the formula proposed in \cite{finite volume} is estimated to be:
    $\Delta m_K^{FV} = -0.22(7) \times 10^{-12}$ MeV.
As for the finite lattice spacing effects, the $\mathcal{O}(a^2)$ error due to the heavy charm is estimated to be the largest source of systematic error. If using physical charm mass and our lattice spacing $a^{-1} = 2.36$ GeV for estimate, this error is relatively $ \sim (m_c a)^2 \sim 25 \%$.
\section{Conclusion and Outlook}
Our preliminary result for $\Delta m_K$ based on 152 configurations with physical quark masses is:
        $$\Delta m_K  = 6.7(0.6)(1.7) \times 10^{-12} MeV.$$
Here the first error is statistical and the second is an estimate of largest systematic error, the discretization error which results from including a heavy charm quark in our calculation. Before making a comparison between our $\Delta m_K$ value and the experimental value $3.483(6) \times 10^{-12}$ MeV, the possibly large finite lattice spacing error needs to be better estimated. We expect the results from our planned $\Delta m_K$ calculations on SUMMIT with a finer lattice spacing will improve the estimate of the systematic errors from discretization effects.


\begin{thebibliography}{99}
\bibitem{bigeng_18}
B. Wang, PoS LATTICE2018, 286 (2018).

\bibitem{gorbahn}
J. Brod and M. Gorbahn, Phys.  Rev.  Lett.  \textbf{108} (2012) , 121801 
\bibitem{jianglei}
N. H. Christ, T. Izubuchi, C. T. Sachrajda, A. Soni and J. Yu, Phys. Rev.  \textbf{D88}(2013), 014508
\bibitem{jianglei_2}
Z. Bai, N. H. Christ, T. Izubuchi, C. T. Sachrajda, A. Soni and J. Yu, Phys. Rev. Lett. \textbf{113}(2014), 112003

\bibitem{xu}
N. H. Christ, X. Feng, A. Jüttner, A. Lawson, A. Portelli, and C. T. Sachrajda, Phys. Rev. \textbf{D94}(2016), 114516
\bibitem{jianglei_pos}
J. Yu, PoS LATTICE2011, 297 (2011).
\bibitem{NPR}
C. Lehner, C. Sturm, Phys. Rev.  \textbf{D84}(2011), 014001
\bibitem{NPR2}
G. Buchalla, A.J. Buras and M.E. Lautenbacher, arXiv:hep-ph/9512380

\bibitem{Sample AMA} 
T. Blum, T. Izubuchi, and E. Shintani, Phys. Rev.  \textbf{D88}(9), 094503 (2013)
\bibitem{finite volume}
N.H. Christ, X. Feng, G. Martinelli and C.T. Sachrajda, arXiv:1504.01170
\bibitem{Bai:2018mdv}
  Z.~Bai, N.~H.~Christ and C.~T.~Sachrajda,
  EPJ Web Conf.\  {\bf 175} (2018) 13017.
  doi:10.1051/epjconf/201817513017
\end{thebibliography}
\end{document}